\documentstyle{article}

\newtheorem{theorem}{Theorem}

\def\ni{\noindent }
\def\eq #1{(\ref{#1})}       
\def\l{\left}                   
\def\r{\right}                  
\def\pa{\partial}               

\newcommand{\be}[1]{\begin{equation}\label{#1}}
\def\ee{\end{equation}}

\newcommand{\ba}[1]{\begin{array}{#1}}
\def\ea{\end{array}}

\def\proof{{\ni\bf Proof. }}

\def\fr #1#2{\frac{#1}{#2}}
\def\frd #1#2{\displaystyle \frac{\displaystyle #1}{\displaystyle #2}}

\def\se #1{sec.\,\ref{#1}}

\def\y1{\mbox{$y'$}}

\def\e{{\rm e}}
\def\F{{F}}
\def\Q{{Q}}
\def\G{{G}}
\def\K{\mathcal K}

\begin{document}
\title{First order ODEs, Symmetries and Linear Transformations}

\author{E.S. Cheb-Terrab$^{a,b}$ and T. Kolokolnikov$^c$ }

\date{}
\maketitle
\thispagestyle{empty}


\centerline {\it $^a$CECM, Department of Mathematics and Statistics}
\centerline {\it Simon Fraser University, Vancouver, Canada.}

\medskip
\centerline {\it $^b$Department of Theoretical Physics}
\centerline {\it State University of Rio de Janeiro, Brazil.}

\medskip
\centerline {\it $^c$Department of Mathematics}
\centerline {\it University of British Columbia,Vancouver, Canada.}

\bigskip
\centerline {\small{(Submitted to {\it ``European Journal of Applied Mathematics"} - July 2000)}}

\maketitle

\begin{abstract}

An algorithm for solving first order ODEs, by systematically determining
symmetries of the form $[\xi=F(x),\, \eta=P(x)\,y+Q(x)]$ - where $\xi\;
\pa/\pa x + \eta\; \pa/\pa y$ is the symmetry generator - is presented. To
these {\it linear} symmetries one can associate an ODE class which
embraces all first order ODEs mappable into separable through linear
transformations $\{t=f(x),\,u=p(x)\,y+q(x)\}$. This single ODE class
includes as members, for instance, 78\% of the 552 solvable first order
examples of Kamke's book. Concerning the solving of this class, a
restriction on the algorithm being presented exists only in the case of
Riccati type ODEs, for which linear symmetries {\it always} exist but the
algorithm will succeed in finding them only partially. 



\end{abstract}

\section{Introduction}

One of the most attractive aspects of Lie's method of symmetries is its
generality: roughly speaking, all solving methods for DEs can be
correlated to particular forms of the symmetry generators
\cite{bluman,stephani}. However, for first order ODEs, Lie's method seems
to be, in principle, not as useful as in the higher order case. The problem
is that the determining PDE - whose solution gives the infinitesimals of
the symmetry group - has the original first order ODE in its
characteristic strip. Hence, finding these infinitesimals requires solving
the original ODE, which in turn is what we want to solve using these
infinitesimals, thus invalidating the approach.

For higher order ODEs, the strategy consists of restricting the cases
handled to the universe of ODEs having {\it point} symmetries, so that the
infinitesimals depend on just two variables, and then the determining PDE
is overdetermined. Although few second or higher order ODEs have point
symmetries, and the solving of such a PDE system for the infinitesimals may
be a major problem in itself \cite{hereman}, the hope is that one will be
able to solve the system by taking advantage of the fact that it is
overdetermined.

One basic motivation in this approach is also that the finite
transformations associated to point symmetries are pointlike and these
transformations form a group {\it by themselves} - not just with respect
to the Lie group parameter. Actually, the composition of {\it any} two
point transformations is also a point transformation. Consequently any
two point symmetries can be obtained from each other through a point
transformation. Lie point symmetries can then be used to tackle the {\it
ODE class} all of whose members can be obtained from (are equivalent to)
each other through point transformations, and which includes a member we
know how to solve (missing the {\it dependent variable}).

Such a powerful approach however is not useful in the case of first order
ODEs - the subject of this paper - for which ``point symmetries" are
already the most general ones. The alternatives left then, roughly
speaking, consist of: looking for particular solutions to the determining
PDE \cite{odetools1}, or restricting the form of the infinitesimals trying
to emulate what is done in the higher order case, such that the problem
can be formulated in terms of an {\it overdetermined} linear PDE
\cite{sym_pat,hydon}. The question in this latter approach, however, is
what would be an ``appropriate restriction" on the symmetries such that:

\begin{itemize}

\item The related invariant ODE family includes a reasonable variety of
non-trivial cases typically arising in mathematical physics;

\item The determination of these symmetries, when they exist, can be
performed systematically, preferably without solving any differential equations;

\item The related finite transformations form a group by themselves
- not just with respect to the Lie group parameter - so that the method
applies to a whole ODE {\it class}.

\end{itemize}

\ni Bearing this in mind, this paper is concerned with first order ODEs
and {\it linear symmetries} of the form

\begin{equation}
\xi=F(x),\ \ \eta = P(x)\, y + Q(x)
\label{sp}
\end{equation}

\ni where $\{\xi,\ \eta\}$ are the infinitesimals, the symmetry generator
is $\xi\, {\pa}/{\pa x} + \eta\, {\pa}/{\pa y}$ and $x$ and $y\equiv y(x)$
are respectively the independent and dependent variables. Concerning the
arbitrary functions $\{F,\,P,\,Q\}$, the requirements are those implied by
the fact that \eq{sp} generates a Lie group of transformations. The linear
symmetries \eq{sp} have interesting features; for instance, the
related finite transformations are also {\it linear}, of the form

\begin{equation}
t=f(x),\ \ u = p(x)\, y + q(x)
\label{tr}
\end{equation}

\ni where $t$ and $u\equiv u(t)$ are respectively the new independent and
dependent variables, and $\{f,\,p,\,q\}$ are arbitrary functions of $x$.
Linear transformations \eq{tr} form a group by themselves too - not just
with respect to the Lie parameter. So, as it happens with point symmetries
in the higher order case, {\it any} two linear symmetries \eq{sp} can be
{\it transformed into each other} by means of a linear transformation
\eq{tr}, and hence to the symmetries \eq{sp} we can associate an {\it ODE
class}. Since separable ODEs have symmetries of this form, the class of
ODEs admitting linear symmetries \eq{sp} actually includes {\it all first
order ODEs which can be mapped into separable} by means of linear
transformations \eq{tr}.



We note that in the particular case of polynomial ODEs, e.g. of Abel type,
\begin{equation}
\y1=f_3(x)\, y^3 + f_2(x)\, y^2 + f_1(x)\, y + f_0(x)
\end{equation}

\ni \eq{tr} actually defines their respective classes\footnote{The Abel
ODE members of a given class can be mapped between themselves through
\eq{tr}. There are infinitely many non-intersecting such Abel ODE classes
but only some of them (still infinitely many) admit symmetries of the form
\eq{sp}.}. Since a separable Abel ODE can be obtained by just taking the
coefficients $f_i$ all equal, this means that there are complete Abel ODE
{\it classes} all of whose members can be transformed into separable by
means of \eq{tr}. Such a case was discussed and solved at the end of the
nineteenth century by Liouville, Appell and others and is presented in
textbooks such as \cite{kamke,murphy}.

More generally, for polynomial ODEs of the form

\begin{equation}
\y1=f_{{n}}(x)\,{y}^{n}+f_{{1}}(x)\, y + f_{{0}}(x)
\label{chini_ode}
\end{equation}

\ni Chini \cite{chini,kamke} presented at the beginning of the twentieth
century a method similar to this mapping into separable but through
transformations \eq{tr} with $q=0$. Chini's method is equivalent to
solving \eq{chini_ode} by determining, when they exist, symmetries \eq{sp}
with $Q=0$.



In connection with the above, this work presents a generalization of these
methods as an algorithm for determining whether or not a first order ODE
of {\it arbitrary} form\footnote{Riccati ODEs are partially excluded from
the discussion - see \se{riccati}.}

$$
\y1=\Psi(x,y),
$$

\ni  belongs to this class admitting
linear symmetries \eq{sp}, and, if so, for explicitly finding the symmetry,
without restrictions to the form of $\{F,\,P,\, Q\}$. Both the
determination of the existence of a solution as well as of the symmetry
itself are performed without solving any auxiliary differential equations.



The exposition is organized as follows. In \se{linear_transformations},
the connection between the symmetries of the form \eq{sp} and linear
transformations of the form \eq{tr} is analyzed and a solving algorithm
for the related ODE class is presented. Some examples illustrating the
type of ODE problem which can be solved using this method are shown in
\se{examples}. In \se{riccati}, a discussion of Riccati ODEs in terms of
their symmetries is made and a variant of the method of
\se{linear_transformations}, to solve a subset of the Riccati problem, is
shown. In \se{tests}, a discussion and some statistics related to the
classification of Kamke's first order ODE examples is presented. Finally,
the conclusions contain general remarks about this work.

\section{Linear transformations and symmetries}

\label{linear_transformations}


To determine whether or not a given first order ODE has symmetries of the
form \eq{sp}, following \cite{sym_pat}, we take advantage of the fact that
for such a symmetry, the related invariant ODE family can be
computed in closed form. With the invariant family in hands we then show how
one can algorithmically compute infinitesimals \eq{sp}, when they exist,
by quadratures; i.e., using only elementary algebraic manipulation,
differentiation, integration and constructing inverses of maps. For
that purpose, with no loss of generality, we first rewrite \eq{sp} in
terms of $\{f(x),p(x),q(x)\}$ entering \eq{tr} as\footnote{In this section
we assume $f' \neq0$, since, otherwise, \eq{GTS} would be a first order
linear ODE. The limiting case where $f' \rightarrow \infty,\, \xi =0$,
however, is not excluded.}

\begin{equation}
\label{sp2}
\xi = \fr{1}{\it f'},\ \ \eta = -{\frac {{\it p'}\,y+{\it q'}}{{\it f'}\,p}}
\end{equation}

%
%

A direct computation of the finite transformations generated by \eq{sp2}
shows that they are linear, of the form
\begin{equation}
t={\rm X},\ \ u={\frac {p(x)\, y+q(x)-q({\rm X})}{p({\rm X})}}
\end{equation}

\ni where ${\rm X}$ is a solution of
$
f({\rm X})-f(x)-\alpha=0
$ and $\alpha$ is the (Lie) group parameter.

Regarding the invariant ODE family related to \eq{sp2}, this family can be
obtained, for instance, by computing differential invariants
$\{I_0,\,I_1\}$ of order 0 and 1 related to \eq{sp2}:

\begin{equation}
I_0=py+q,\ \ I_1={\frac {{\it f'}}{{\it p'}\, y + p \y1 +{\it q'}}}
\end{equation}

\ni and hence, the invariant ODE, given by $\Lambda(I_0,I_1) = 0$, where
$\Lambda$ is arbitrary, can be conveniently written as $I_1 \G(I_0) = 1$,
with arbitrary $\G$, resulting in

\begin{equation}
\label{GTS}
\y1=\fr{f'}{p}\G(py+q) - \fr{q'}{p} - \fr{p'}{p} y
\end{equation}

\ni Due to this connection between linear transformations and symmetries
\eq{sp}, the same invariant family \eq{GTS} can be obtained
directly from an autonomous ODE,

\begin{equation}
\label{auto}
{u'}=\G(u)
\end{equation}

\ni by just changing variables in it using \eq{tr}. The solution to
\eq{GTS} can be obtained in the same way, by changing variables in the
solution to \eq{auto}:

\begin{equation}
\label{GTS sol}
f-\int ^{py+q}\! \fr{1}{\G(z)}{dz}+C_1=0
\end{equation}

\ni In fact \eq{sp2} too can be obtained from the symmetry
$[\xi=1,\,\eta=0]$ of \eq{auto} by changing variables using \eq{tr}. We
recall that the knowledge of $\{\xi,\,\eta\}$ entering \eq{sp2} suffices
to express the solution of any member of the class \eq{GTS} by
quadratures and so it is equivalent to the determination of
$\{f,\,p,\,q\}$ entering \eq{GTS sol}.

\begin{theorem} \label{thm GTS}

Consider an ODE $y'=\Psi(x,y)$ with $\Psi_{yyy} \ne 0$. Both determining
whether or not this ODE belongs to the class \eq{GTS}, and, in the
positive case, computing the infinitesimals \eq{sp2} themselves, can be
performed algorithmically from $\Psi$ by quadratures.

\end{theorem}


\proof We start by noting an intrinsic feature of ODEs that are members of
the class \eq{GTS}: the determination of $p(x)$ up to a constant factor -
say $\kappa$ - suffices to map any member of the class into another one
having a symmetry of the form

\begin{equation}
\label{sp3}
\xi=\F(x),\ \ \eta=\Q(x)
\label{X_FxQx}
\end{equation}

\ni This is easily verified by changing variables
\begin{equation}
\label{tr_p}
y=\fr{u}{\kappa\,p}\
\end{equation}

\ni in \eq{sp2}, arriving at $[\xi=1/{f'},\, \eta=-{{q'\kappa}/{f'}}]$,
which is of the form \eq{sp3}. In turn, symmetries of the form \eq{sp3},
when they exist, can be systematically determined as shown in
\cite{sym_pat}. In what follows we develop the proof by first showing how
to map any ODE member of \eq{GTS} into one having a symmetry of the form
\eq{sp3}, and then, for completeness, briefly reviewing how such a
symmetry is determined when it exists.


Now, in view of \eq{GTS}, for all ODEs $\y1=\Psi$ members of that class we have that

\begin{equation}
\ba{lll} 
\Psi_y &=& f' \G'(p y+q) - \frac{p'}{p},
\\*[.1in]
\Psi_{yy} &=& f' p \G''(p y+q),
\\*[.1in]
\Psi_{yyy} &=& f' p^2 \G'''(p y+q).\\
\ea
\end{equation}

\ni Thus let
\begin{equation}
\label{G2/G3}
A\equiv\frd{\Psi_{yy}}{\Psi_{yyy}}=\frac{1}{p} K(p y + q)
\ \ \mbox{ where }\ \  K=\frac{\G''}{\G'''}.
\end{equation}

\ni Three cases now arise, respectively related to whether $A_y=0$,
$A_{yy}=0$ or $A_{yy} \neq 0$.

\subsubsection*{\underline{Case $A_y = 0$}}

In this case, $K'=0$, so that $K = \kappa$ for some non-zero constant
$\kappa$, and hence $A={\kappa}/{p}$.
So, from \eq{tr_p}, when $\y1=\Psi$ is indeed a member of this class, by
changing variables using

\begin{equation}
\label{tr_Ay_is_zero}
y=A\, u
\end{equation}

\ni the resulting ODE family will have a symmetry of the form \eq{sp3}.

\subsubsection*{\underline{Case $A_{yy} = 0$, $A_{y} \neq 0$}}

In this case, $K''=0$ so that from \eq{G2/G3}
\begin{equation}
A =
\frac{\kappa_1\, (p\,y+q)+\kappa_0}{p}
\end{equation}

\ni for some constant $\kappa_0$ and some non-zero constant $\kappa_1$.
Here the necessary condition for $\y1=\Psi$ to be a member of \eq{GTS} is
that the ratio above be linear in $y$. In such a case, when the ODE
is indeed a member of this class, by introducing 

\begin{equation}
\label{tr_Ayy_is_zero}
u = \ln(A)
\end{equation}

\ni the resulting ODE in $u$ will have a symmetry of the form \eq{sp3}; this
can be verified straightforwardly by performing the change of variables
directly in \eq{sp2}.

\subsubsection*{\underline{Case $A_{yy} \neq 0$}}
In this case let 
\begin{equation}
\label{I}
I \equiv \fr{A_{yx}}{A_{yy}} = \frac{p' y + q'}{p}
\end{equation}

\ni The necessary condition for $\y1=\Psi$ to be a member of
\eq{GTS} is then that $I$ be linear in $y$,
from where
\begin{equation}
\label{p}
p(x) = \exp\left( \int I_y dx\right)
\end{equation}

\ni So, when the ODE belongs to this class, from \eq{tr_p},
changing variables $y=u/p$ will lead to an ODE having a symmetry of the form \eq{sp3}.

Once we have shown how a member of \eq{GTS} can be mapped into another one
having a symmetry of the form \eq{sp3}, what remains to be done in the proof of
Theorem \ref{thm GTS} is to review how that symmetry can be
obtained by quadratures.


\subsection{Symmetries of the form $[\xi=\F(x),\ \eta=\Q(x)]$}
\label{FxQx}

By computing differential invariants, as we did to arrive at \eq{GTS}, the
invariant ODE family associated to $[\xi=\F(x),\ \eta=\Q(x)]$ can be
written as\footnote{If $\Q=0$ or $\F=0$ then the invariant ODE is
separable or linear, so in \eq{ODE_pat_FH} and henceforth we assume $\F \neq
0,\, \Q \neq 0$. }:

\begin{equation}
\y1= \Phi(x,y) \equiv 
\frac{1}{\F(x)}
\left(\Q(x)+\G\l(y-\int \!{\frac {\Q(x)}{\F(x)}}{dx}\r)\right)
\label{ODE_pat_FH}
\end{equation}

\ni where $\F$, $\Q$ and $\G$ are arbitrary functions of their arguments.
So far we have shown that if an ODE belongs to \eq{GTS}, then after
changing variables as shown for the cases $A_y =0,\, A_{yy}=0,\,
A_{yy}\neq 0$, the resulting ODE will be a member of this family
\eq{ODE_pat_FH}.

Now, to determine $\F$ and $\Q$, following \cite{sym_pat}, we first build
an expression depending on $x$ and $y$ only through $\G$

\begin{equation}
\K \equiv \frac{\Phi_y}{\Phi_{yy}}
=
\frac{\G_y}{\G_{yy}}
\label{Q_eq}
\end{equation}

\ni where we assume\footnote{If $\Phi_{yy} = 0$, then \eq{ODE_pat_FH} is
already a first order linear ODE solvable in terms of
quadratures.} $\Phi_{yy} \neq 0$. As explained in \cite{sym_pat},
the problem then splits into two cases.

\subsubsection*{\underline{Case $\K_{y} \neq 0$}}

In this case, we can obtain the ratio $\Q(x)/\F(x)$ - only depending on
$x$ - by taking

\begin{equation}
\Upsilon \equiv \frac{\K_x}{\K_y}
=
-\frac{\Q(x)}{\F(x)}
\label{ratio_HF}
\end{equation}

\ni The knowledge of this ratio in turn permits the elimination of $\Q$
from the determining PDE for the infinitesimals, leading to\footnote{For
more details see \cite{sym_pat}}

\begin{equation}
\label{F}
\F(x)=C_1\,{\e^{^{
\displaystyle
\int \!{\l(
\frac {\Upsilon\,\Phi_{{y}} - \Upsilon_{{x}}-\Phi_{{x}}}{\Phi +
\Upsilon}\r)}\ {dx}}}}
\end{equation}

\ni which together with \eq{ratio_HF} gives the solution we
are looking for. The necessary and sufficient conditions for the existence
of such a symmetry are:

\begin{equation}
\frac{\pa}{\pa y}
\l(\frac{\K_x}{\K_y}\r)=0,
\ \ \ \
\frac{\pa}{\pa y}
\l(
\frac {\Upsilon\,\Phi_{{y}} - \Upsilon_{{x}}-\Phi_{{x}}}{\Phi +
\Upsilon}\r)=0
\label{diff_conditions}
\end{equation}

\subsubsection*{\underline{Case $\K_{y} = 0$}}

Since $\K_y=-\K_x\,\F/\Q$, then when $\K_y=0$, $\K_x=0$ too, so
$\K=\kappa$ for some non-zero constant $\kappa$. Hence, the
right-hand-side of \eq{ODE_pat_FH} satisfies:

\begin{equation}
\frac{\Phi_y}{\Phi_{yy}}=\kappa
\end{equation}

\ni and so \eq{ODE_pat_FH} - the invariant ODE family - is of the form

\begin{equation}
\y1 = \Phi \equiv A(x)+B(x)\ 
{\e^{^{\displaystyle{{y}/{\kappa}}}}}
\label{pat_2}
\end{equation}

\ni where $A$ and $B$ are arbitrary functions. For a given ODE of this type,
$A$ and $B$ can be determined by inspection, and the determining PDE for the
infinitesimals can be solved directly in terms of $A$ and $B$ as
\begin{equation}
\F(x)  = 
{\frac
{\e^{^{\displaystyle-\int \!{\frac {A}{\kappa}}dx}}}{B}},
\ \ \ \
\Q(x) = A\, \F(x)
\end{equation}

\subsection{Examples}
\label{examples}

1. Consider the first order ODE example number 128 from Kamke's book:

%
%

\begin{equation}
\label{k128}
x\y1+ay-f(x)g({x}^{a}y)=0
\end{equation}

\ni For this ODE, Kamke shows a change of variables mapping the ODE into
separable, derived for this particular ODE family in \cite{dickson}. Using
the algorithm presented in this paper, we tackle this ODE by computing $A$
in \eq{G2/G3}

\begin{equation}
\label{A_k128}
A={\frac {{\it g''}}{{x}^{a}{\it g'''}}}
\end{equation}

\ni so we are in {\it ``Case $A_{yy} \neq 0$"}. We then proceed with
computing $I$ (see \eq{I}) arriving at

\begin{equation}
\label{I_k128}
I = \fr{a\,y}{x}
\end{equation}

\ni The existence condition that $I$ be linear in $y$ is satisfied,
hence, according to \eq{p},
\begin{equation}
p(x) = x^a
\end{equation}

\ni Now, changing variables $y=\frd{u}{p}$ as indicated in
\eq{tr_p}, \eq{k128} becomes

\begin{equation}
\label{k128_xx}
u'=g(u)\,{\frac {f(x){x}^{a}}{x}}
\end{equation}

\ni which is already {\it separable} (and thus naturally has a symmetry of
the form \eq{sp3}). The solution to \eq{k128} is then obtained by changing
variables back in the solution to \eq{k128_xx}, leading to:
\begin{equation}
\label{ex_2}
\int \! {x}^{a-1}\,f(x)\,{dx}-\int ^{{x}^{a}y}\!\fr{1}{g(z)}{d
z}-C_1=0
\end{equation}

2. Let's now discuss an example for which $A_{yy}=0$,


\begin{equation}
\label{ex_3}
\y1=\left ({x}^{3}{y}^{4} + 4\,{x}^{4}{y}^{3}+6\,{x}^{5}{y}^{2} + 4\,{x}^{6}y + {x}^{7}\right )\left ({x}^{n}+1\right )-{\frac {y}{x}}-2
\end{equation}

\ni where $n$ is an arbitrary constant. For this ODE, from \eq{G2/G3},
\begin{equation}
A = \frac{y+x}{2}
\end{equation}

\ni so the change of variables here is $u=\ln(A)$ (see \eq{tr_Ayy_is_zero}), mapping \eq{ex_3} into
\begin{equation}
\label{ex_3_u}
u'={t}^{7+n}\left ({t}^{n}+1\right ){e^{6\,u}}+\fr{7}{t}
\end{equation}

\ni For this ODE, a symmetry of the form \eq{sp3} is computed algorithmically
(see \se{FxQx})
\begin{equation}
\xi=\fr{1}{8\,\left ({t}^{n}+1\right )},\ \
\eta=-{\frac {1}{8\,t\left ({t}^{n}+1\right )}}
\end{equation}

\ni Changing variables back directly in the above we arrive at a symmetry for \eq{ex_3}
\begin{equation}
\xi= \fr{1}{8\,\left ({x}^{n}+1\right )}, \ \ 
\eta=-{\frac {y + 2\,x}{8\,x\left ({x}^{n}+1\right )}}
\end{equation}

\ni from where an implicit solution to \eq{ex_3} follows as 
\begin{equation}
x + {\frac {1}{3\,{x}^{3}\left (y+x\right )^{3}}}+{\frac {{x}^{(1+n)}}{1+n}}=C_1
\end{equation}

3. As an example of the case in which $A_y=0$, consider

\begin{equation}
\label{ex2}
\y1= {b\,e^{a\,x\,y}}{x}^{a}
    + {\frac {\left ({x}^{2}-1\right )y}{x}}
    - \fr{1}{x^{2}}
    + \ln (x)+c
\end{equation}

\ni where $a$, $b$ and $c$ are arbitrary constants. From \eq{G2/G3}, $A = 1/(a\,x)$, so that by changing variables
as indicated in \eq{tr_Ay_is_zero}, \eq{ex_2} becomes
\begin{equation}
\label{ex2_2}
u'=ut+a\,b\,{t}^{(a+1)}{e^{u}}-{\frac {a}{t}}+a\,t\l(\ln (t)+c\r)
\end{equation}

\ni and this ODE has a symmetry of the form \eq{sp3}

\begin{equation}
\xi = \fr{1}{t},\ \eta = -\fr{a}{t^2}
\end{equation}

\ni from where, by changing variables back, \eq{ex2} admits the symmetry
\begin{equation}
\xi=\fr{1}{x}, \ \ 
\eta=-{\frac {x\,y+1}{{x}^{3}}}
\end{equation}

\ni which suffices to integrate \eq{ex2} by either using canonical
coordinates or computing an integrating factor.

4. The algorithm presented is applicable to higher degree ODEs too (see
Table 1. in \se{tests}), provided that it is possible to solve 
the given ODE for $y'$. Consider for instance Kamke's example 394

\begin{equation}
({\y1})^{2}+2\,f\,y\,\y1+g\,{y}^{2}+\left ({f}^{2}-g\right ){\e^{-2\,\int _{a}^
{x}\!f(z){dz}}}=0
\label{ex_4}
\end{equation}

\ni where $f\equiv f(x)$ and $g\equiv g(x)$ are arbitrary functions. For
this problem Kamke presents a particular change of variables derived in
\cite{pirondini}. To tackle this example using our algorithm we first
solve the ODE for $y'$

\begin{equation}
\y1=-fy \pm \sqrt {\left ({y}^{2}-{\e^{-2\,\int _{a}^{x}\!f(z){dz}}}\right 
)\left ({f}^{2}-g\right )}
\label{hd}
\end{equation}

\ni Now, by taking any of the two branches of \eq{hd}, for instance the ``+" one,
and computing the second derivative of \eq{G2/G3}, we find
\begin{equation}
A_{yy} = {\frac {2}{3\,{y}^{3}\left ({\e^{\int _{a}^{x}\!f(z){dz}}}\right )^{2}}}
\end{equation}

\ni so that $I$ in \eq{I} is given by
\begin{equation}
I = f\,y
\end{equation}

\ni from where we compute $p$ using \eq{p}, finally arriving at the symmetry

\begin{equation}
\xi={\frac {1}{\sqrt {{f}^{2}-g}}},\ \ 
\eta=-{\frac {f\,y}{\sqrt {{f}^{2}-g}}}
\end{equation}

\ni actually admitted by both branches of \eq{hd}.

It is worth mentioning that if on one hand these four examples are
straightforward problems for the single solving algorithm presented, on
the other hand we are not aware of any other algorithm for tackling
examples like 2 or 3; also, for examples 1 and 4 the changes of variables
presented in Kamke are non-obvious and presented in connection with {\it
different} problems \cite{dickson,pirondini}. Despite the presence of
arbitrary functions and parameters, both Kamke's examples 128 and 394 are
actually particular cases of the class represented by \eq{GTS}.

\section{Riccati equations}
\label{riccati}

The case of Riccati type ODEs
\begin{equation}
\y1=f_{{2}}\,{y}^{2}+f_{{1}}\,y+f_{{0}}
\label{ric_ode}
\end{equation}

\ni where $f_i \equiv f_i(x)$, $f_2 \neq 0$ and $f_0 \neq 0$, deserves a
separate discussion. All Riccati ODEs admit symmetries of the form
\eq{sp} and so all of them can be mapped into separable using
transformations of the form \eq{tr}. However, it is easy to verify that to
find such a transformation requires solving the Riccati ODE
itself. The algorithm of the previous section - which does not rely in
solving auxiliary differential equations - only works when
$\Psi_{yyy}\neq0$ - see Theorem 1. The usual approach for solving
\eq{ric_ode} then consists of converting it to a linear second order ODE
and using the various methods available for this other problem.

Nonetheless, there are entire subfamilies of \eq{ric_ode} for which
symmetries of the form \eq{sp} can be found following an approach such as
the one presented in the previous section, without using techniques for
linear second order ODEs. Such an approach is interesting since it
enriches the algorithms available for tackling the problem and could be of
use for solving some linear ODEs by mapping them into Riccati ones as
well. For the purpose of discussing these cases and without loss of
generality we rewrite \eq{sp2} - the general form of the symmetries
admitted by Riccati ODEs - redefining $f' \rightarrow f/p$

\begin{equation}
\xi = {\frac {p}{f}},\ \eta = -{\frac {{\it p'}\,y+{\it q'}}{f}}
\label{sp4}
\end{equation}

\ni If now, in \eq{GTS}, we redefine $f'$ in the same way and take $G$ as a
square mapping depending on two arbitrary constants $a$ and $b$,

\begin{equation}
G=u\mapsto {u}^{2}+a\,u+b
\end{equation}

\ni we arrive at the form of an arbitrary Riccati ODE - as general as
\eq{ric_ode} - but expressed in terms of these two constants $a$ and $b$
and the functions $\{f,\,p,\,q\}$ appearing in its symmetry generator
\eq{sp4}:

\begin{equation}
\y1=f\,{y}^{2}+{\frac {\left (a+2\,q\right )f-{\it p'}}{p}}\,y
+ {\frac {\left (\left (a+q\right )q+b\right )f-{\it q'}\,p}{{p}^{2}}}
\label{ric_ode_fpq}
\end{equation}

\smallskip
\ni{\underline{Case $p' = 0$}}
\smallskip

\ni A first solvable case happens when $p'=0$, so that in \eq{sp4} both
infinitesimals depend only on $x$ and hence the symmetry can be
systematically determined - also for Riccati ODEs - as explained in
\se{FxQx}.

\smallskip
\ni{\underline{Case $q' = 0$}}
\smallskip

A second solvable case happens when, in \eq{ric_ode_fpq}, $q'=0$, so that
the infinitesimals \eq{sp4} are of the form
\begin{equation}
\xi={\mathcal F}(x),\ \eta={\mathcal P}(x)\,y
\label{chini_symmetry}
\end{equation}

\ni An algorithm for solving such an ODE was presented by Chini
\cite{chini}. In the case of Riccati ODEs, Chini's algorithm can be
summarized as {\it ``to check for the constant character"} of the
expression\footnote{This connection between the constant character of an
expression like \eq{chini_invariant}, built with the coefficients of a
polynomial ODE of the form \eq{chini_ode}, and symmetries of the form
\eq{chini_symmetry}, is valid not only for Riccati ODEs but for ODEs of
the form \eq{chini_ode} in general.}

\begin{equation}
{\mathcal I} \equiv {\frac {\left ({f_{{0}}}^{'}f_{{2}}-f_{{0}}{f_{{2}}}^{'}-2\,f_{{0}}f_{{1}}f_{{2}}\right )^{2}}
{\left (f_{{0}}f_{{2}}\right )^{3}}}
\label{chini_invariant}
\end{equation}

\ni where $f_i$ are the coefficients of $y$ in \eq{ric_ode}. Whenever
${\mathcal I}$ is constant, the problem is systematically solvable in
terms of quadratures (see for instance \cite{kamke} - p.303). Concerning
\eq{ric_ode_fpq} at $q'=0$, a direct computation of ${\mathcal I}$
confirms that in such a case ${\mathcal I}$ is constant. Conversely,
another direct computation shows that {\it whenever ${\mathcal I}$ is
constant}, the ODE will have a symmetry of the form \eq{chini_symmetry}.
To check that, it suffices to solve \eq{chini_invariant} for $f_1$ and
substitute the result into \eq{ric_ode}; the resulting Riccati ODE will admit the
symmetry

\begin{equation}
\xi = \fr{1}{f_{{2}}}\sqrt {{\frac {f_{{2}}}{f_{{0}}}}},\ \ 
\eta =
\fr{\left ({f_{{0}}}^{'}f_{{2}} - f_{{0}} {f_{{2}}}^{'}\right )}
{2\,{f_{{0}}}^{2}{f_{{2}}}{\sqrt {{\frac {f_{{2}}}{f_{{0}}}}}}}\, y
\end{equation}

\ni which is of the form \eq{chini_symmetry}.

We can also see what is the ODE class solved by Chini's algorithm, as well
as explain the previous results, by noticing that \eq{chini_invariant} is
an absolute invariant for \eq{ric_ode} under transformations of the form

\begin{equation}
t = {\tilde f}(x),\ \ u={\tilde p}(x)\,y\
\label{chini_tr}
\end{equation}
that is, of the form \eq{tr} with $q=0$. In fact, \eq{chini_invariant} can be written as 

\begin{equation}
{\mathcal I} = \fr{{s_3}^2}{{s_2}^3}
\label{I_s2_s3}
\end{equation}
\ni where
\begin{equation}
s_2  =  f_{{0}}f_{{2}},\ \ \ \ \ \ 
s_3  =  {f_{{0}}}^{'}f_{{2}}-{f_{{2}}}^{'}f_{{0}}-2\,f_{{0}}f_{{1}}f_{{2}},
\label{invariants}
\end{equation}
\ni are relative invariants of weight 2 and 3 with respect to
transformations \eq{chini_tr} \cite{ktese}.

In summary: Chini's algorithm solves the ODE class generated by changing
variables \eq{chini_tr} in the general Riccati ODE \eq{ric_ode_fpq} at
$q'=0$, all of whose members have ${\mathcal I} = \mbox{\it constant}$.

Three additional solvable Riccati families, where the invariant $\mathcal
I$ is {\it non-constant}, are obtained by equating in \eq{sp4} any two of
the three arbitrary functions $\{f,\,p,\,q\}$ with $p' \neq 0$ and $q'
\neq 0$.

\smallskip
\ni{\underline{Case $f = p$}}
\smallskip

When a given Riccati ODE belongs to this family, then by changing
variables\footnote{$f$ is the coefficient of $y^2$ in the given ODE.} 

\begin{equation}
y={\frac {u}{f}}
\label{tr_f=p}
\end{equation}

\ni in the given ODE and in the general form of its symmetry
\eq{sp4}, we see that the resulting ODE in $u$ will admit the symmetry

\begin{equation}
\xi=1,\ \eta=-{\it q'}
\end{equation}
\ni which can be determined as explained in \se{FxQx}.

\smallskip
\ni {\underline{Case $q = p$}}
\smallskip

When a given Riccati ODE belongs to this family, then by changing variables

\begin{equation}
y=u-1
\label{tr_q=p}
\end{equation}

\ni in the given ODE and in \eq{sp4} we see that 
the resulting ODE in $u$ will admit a symmetry of the form

\begin{equation}
\xi={\frac {p}{f}},\ \eta=-{\frac {{\it p'}}{f}}\,u
\end{equation}
\ni that is, infinitesimals of the form \eq{chini_symmetry}, and hence the
ODE will be solvable using Chini's method.

\smallskip
\ni {\underline{Case $f = q$}}
\smallskip

From \eq{ric_ode_fpq}, the Riccati family corresponding to
this case is given by
\begin{equation}
\y1=f{y}^{2}
+ {\frac {\left (f\left (a+2\,f\right )-{\it p'}\right )}{p}}\,y
+ {\frac {\left ((a+{f})f+b\right )f-{\it f'}\,p}{{p}^{2}}}
\label{ric_f=q}
\end{equation}

\ni From \eq{sp4}, this ODE family admits the symmetry 
\begin{equation}
\xi={\frac {p}{f}},\ \eta=-{\frac {{\it p'}\,y+{\it f'}}{f}}
\label{sp5}
\end{equation}

\ni We haven't found an obvious transformation of the form $y=P\,u+Q$ to map this
symmetry to one of the forms \eq{X_FxQx} or \eq{chini_symmetry}. A possible
approach would then be to directly set up the determining PDE 

$$
\eta_{{x}}+\left (\eta_{{y}}-{\xi}^{'}\right )\left (f_{{2}}{y}^{2}+f_
{{1}}y+f_{{0}}\right )-\xi\,\left ({f_{{2}}}^{'}{y}^{2}+{f_{{1}}}^{'}y
+{f_{{0}}}^{'}\right )-\eta\,\left (2\,f_{{2}}y+f_{{1}}\right ) = 0
$$

\ni for the coefficients $\xi$ and $\eta$ of the infinitesimal symmetry
generator $\xi \frac{\partial}{\partial x} + \eta \frac{\partial}{\partial
y}$ of an arbitrary Riccati ODE \eq{ric_ode}. Then take $\xi$ and $\eta$
as given by \eq{sp5} and run a differential elimination process solving
for $p$. That approach works but the resulting symbolic expressions are
large enough to become untractable even with simple examples.

An alternative approach leading to more tractable expressions is based on
using the information we have in \eq{ric_f=q} concerning the existence of
two constants $a$ and $b$. So, by equating the coefficients of
\eq{ric_f=q} with those of \eq{ric_ode} we arrive at the system

\begin{eqnarray}
\lefteqn{f_2 - f = 0,} & &
\label{sys_pab}
\\
&  & 
f_{{1}}-{\frac {f_{{2}}\left (a+2\,f_{{2}}\right )-{\it p'}}{p}}=0,\ \ \ 
f_{{0}}-{\frac {f_{{2}}\left (f_{{2}}a+{f_{{2}}}^{2}+b\right )-{f_{{2}}}^{'}p}{{p}^{2}}}=0
\nonumber
\end{eqnarray}

\ni This system can be solved for $a$ and $b$, so that when a Riccati
equation belongs to this family the following two expressions formed
with its coefficients $f_i$ will be constants:

\begin{eqnarray}
a & = & {\frac {f_{{1}}\,p-2\,{f_{{2}}}^{2}+{\it p'}}{f_{{2}}}}
\nonumber
\\ 
b & = & {\frac {f_{{0}}\,{p}^{2} + f_{{2}}\left ({f_{{2}}}^{2} - f_{{1}}p - {\it p'} \right )
+ {f_{{2}}}^{'}p}{f_{{2}}}}
\label{sol_b}
\end{eqnarray}

\ni At this point, however, we cannot verify the constant character of
these expressions because $p$ is still unknown. An expression for $p$ can
be obtained by computing the integrability conditions for $f_2$ and $p$
implied by \eq{sys_pab}:

\begin{eqnarray}
{f_{{2}}}^{{\it ''}} & = & f_{{1}}{f_{{2}}}^{'}-2\,f_{{0}}{\it p'}-p{f_{{0}}}^{'}
+ {\frac {\left ({f_{{2}}}^{'}\right )^{2}+pf_{{0}}{f_{{2}}}^{'}}{f_{{2}}}},
\label{eq_p'}
\\
{\it p''} & = & 2\,{f_{{2}}}^{'}f_{{2}}-{\it p'}
\,f_{{1}}-p{f_{{1}}}^{'}+{\frac {{f_{{2}}}^{'}{\it p'}+pf_{{1}}{f_{{2}}}^{'}}{f_{{2}}}}
\label{eq_p''}
\end{eqnarray}

\ni Using \eq{eq_p'} to eliminate $p'$ from \eq{eq_p''} we
arrive at a solution for $p$:
\begin{eqnarray}
\lefteqn{p=} & &
\label{ans_p}
\\*[.11in]
& & {\frac {f_{{2}}\left (
  3s_2\,{f_{{2}}}^{{\it ''}}{s_2}^{'}
- 2{s_2}^{2}{f_{{2}}}^{{\it '''}}
+ \left (
    ({s_2}^{{\it ''}}-s_4)\,s_2
    -8{s_2}^{3}
    - 2\left ({s_2}^{'}\right )^{2}
    + 2{s_3}^{2}
    \right ){f_{{2}}}^{'}\right )}
    {s_{{2}}\left (
        2\,s_{{2}}{s_{{3}}}^{'}
        - 3\,s_{{3}}{s_{{2}}}^{'}
        \right )}
}
\nonumber
\end{eqnarray}
\ni where
\begin{equation}
s_4  = 
{\frac
    {2\,s_2\,{s_3}^{'}-3\,s_3\,{s_2}^{'}+3\,{s_3}^{2}}
    {2\,s_2}}
\label{invariant_s4}
\end{equation}

\ni is the relative invariant of weight 4 for Riccati ODEs with respect to
transformations of the form \eq{chini_tr} \cite{ktese}. We note that, from
\eq{I_s2_s3}, when the denominator of \eq{ans_p} is zero the ODE has a
{\it constant} invariant ${\mathcal I}$ \eq{chini_invariant} and so it is
already solvable using Chini's algorithm.

In summary, a strategy - not relying on solving second order linear ODEs -
for finding linear symmetries of the form \eq{sp} for Riccati equations could
consist of

\begin{enumerate}

\item Check if the ODE has a symmetry of the form $[\xi={\mathcal
F}(x),\,\eta={\mathcal Q}(x)]$ (algorithm of \cite{sym_pat} - see \se{FxQx}); or
$[\xi={\mathcal F}(x),\,\eta={\mathcal P}(x)\,u]$ (Chini's algorithm);

\item Check if the ODE belongs to one of the two families ``$f=p$" or
``$p=q$" by using the transformations \eq{tr_f=p} and \eq{tr_q=p} and re-entering
the previous step;

\item Check if the ODE belongs to the family ``$f=q$" \eq{ric_f=q}; for that purpose:
\begin{enumerate}

\item compute the invariants \eq{invariants};

\item use these invariants to compute $p$ using \eq{ans_p};

\item plug the resulting $p$ into \eq{sol_b} and verify if
the two right-hand-sides are constant - if so, the ODE admits the symmetry
\eq{sp5}

\end{enumerate}
\end{enumerate}

It is our belief that with the development of computer algebra software
and faster computers, these type of algorithms for Riccati subclasses
(there are of course many other possibilities) will become each day more
relevant, also as an alternative for tackling second order linear ODEs.

\section{Classification of Kamke's examples and discussion}
\label{tests}

We have prepared a computer algebra prototype of the algorithms presented
in \se{linear_transformations} and \se{riccati} using the Maple system. We
then used this prototype to analyze the set of Kamke's 576 first order ODE
examples. This type of computational activity interesting. In the first
place, Kamke's book contains a fair collection of examples arising in
applications. In the second place, this book collects many, perhaps most,
of the solving algorithms available in the literature for first order
ODEs. So, generally speaking, such an analysis conveys a reasonable
evaluation of the range of application and novelty of a new ODE solving
algorithm.

In order to perform the classification, we first excluded from the 576
Kamke examples all those for which a solution is not shown and we were not
able to determine it by other means\footnote{The numbers of the Kamke
examples we excluded in this way are: 47, 48, 50, 55, 56, 74, 79, 82, 202,
205, 206, 219, 234, 235, 237, 265, 250, 253, 269, 331, 370, 461, 503 and
576.} - most of them just because the ODE is too general. So our testing
arena starts with 552 ODEs.

Then, we know that all ODEs of type separable, linear, homogeneous,
Bernoulli, Riccati and Abel with constant invariant\footnote{For an
enumeration of Kamke's examples of Abel type with constant invariant see
\cite{abel} and concerning other classes see \cite{odetools1}.}, that is,
372 of Kamke's examples, have symmetries of the form \eq{sp} and then
belong to the ODE class discussed in this paper. So the first thing we
wanted to know is how many of the remaining 552 - 372 = 180 examples admit
linear symmetries of the form \eq{sp}? Is there any other classification
known for these 180 examples? The information for answering these and
related questions is summarized in this table:

\begin{center}
{
\begin{tabular}{lccc}
\hline
\hline
Class & First degree in $y'$: 88 & Higher degree in $y'$: 92 & Total: 180 ODEs \\
\hline
$[\xi=F,\ \eta=Py+Q]$
& 20 & 37 & 57 \\
\hline
Abel (non-constant invariant) & 15 & 0 & 15 \\
\hline
Clairaut & 0 & 15 & 15  \\
\hline
d'Alembert & 2 & 21 & 23 \\ 
\hline
{\it Unknown} & 33 & 21 & 54 \\ 
\hline
\hline
\multicolumn{4}{c}{Table 1. Classification of 180 {\it non} ``linear, separable,
Bernoulli, Riccati or Abel c.i." Kamke's examples.} 
\end{tabular}
}
\end{center}
\bigskip

From these numbers, some first conclusions can be drawn. First, 372 + 57 =
429 ODEs, that is, 78 \% of Kamke's 552 solvable examples, have linear
symmetries of the form \eq{sp}. Also Table 1. shows that in Kamke - even among these
particular 180 ODEs which exclude the easy ones - there are more examples
having symmetries of the form \eq{sp} which can be systematically
determined as explained in \se{linear_transformations}, than examples of
Abel (non-constant invariant), Clairaut and d'Alembert types
all together\footnote{We note there is no intersection between these
classifications: all of Abel (non-constant invariant) Clairaut and
d'Alembert equations do not have symmetries of the form \eq{sp}.}.

In the second place, if we discard the 61 examples of Riccati type found in
Kamke, the solvable set is reduced to 552 - 61 = 491 ODEs, and from this
set, 429 - 61 = 368 ODEs, that is, 75 \% of these 491 can be solved
algorithmically as shown in \se{linear_transformations}.

Moreover, a classification of Kamke's examples of Riccati type according to
\se{riccati} shows:


\begin{center}
{
\begin{tabular}{lcccc}
\hline
\hline
Class & $[\xi=F,\ \eta=Q]$ & $[\xi=F,\ \eta=Py]$ & ``Two of $\{f,p,q\}$ are equal" & Total of ODEs \\
\hline
Riccati & 7/61 & 22/61 & 2/61 & 31/61 \\
\hline
\hline
\multicolumn{5}{c}{Table 2. Classification according to \se{riccati} of the 61 Riccati ODE examples of Kamke's book.} 
\end{tabular}
}
\end{center}
\bigskip

\ni So one half of these Riccati examples is still solvable using the
algorithms described in \se{riccati}, and so without mapping the problem
into a second order linear ODE nor having to solve auxiliary differential
equations.




\section{Conclusions}

In this work we presented an algorithm for solving first order ODEs,
consisting of determining symmetries of the form \eq{sp}. From the discussions of
\se{linear_transformations}, to these symmetries one can associate an ODE
class - represented by \eq{GTS} - which embraces all first order ODEs
mappable into separable ones through linear transformations. From the
numbers of \se{tests}, this class appears to us as the widest {\it first
order ODE class} we are aware of, all of whose members are algorithmically
solvable as shown in \se{linear_transformations} or \se{riccati}, or
mappable into second order linear ODEs when the ODE is of Riccati type
and the methods in \se{riccati} don't cover the case\footnote{In fact the
solving of the equivalence problem for Riccati ODEs with respect to linear
transformations is equivalent to solving the whole Riccati class; in turn
a too general problem.}.

The algorithms presented also do not require solving additional
differential equations nor do they rely on the ODE member of the class being
{\it algebraic} (i.e.: rational in $y$ and its derivatives) or on
restrictions to the function fields - the only requirement on the
functions entering the infinitesimals \eq{sp} are those implied by the
fact that these infinitesimals do generate a Lie group of transformations.

Concerning other related works we are aware of, the method presented in
\cite{schwarz} for solving Abel ODEs with {\it constant} invariant is a
particular case of the one presented here in that those ODEs are the
subclass of \eq{GTS} of type Abel. In the same line, the method by Chini
\cite{chini} - a generalization of the method for Abel ODEs with constant
invariant which solves more general ODEs of the form \eq{chini_ode} - is
also a particular case in that \eq{chini_ode} is a very restricted
subclass of \eq{GTS}. Also, the class solvable through Chini's method is
equivalent to a separable ODE only through transformations \eq{tr} with
$q=0$. In this sense, the algorithm presented in
\se{linear_transformations} generalizes both the one discussed in
\cite{schwarz} and the one presented in \cite{chini}.

The fact that this class \eq{GTS} is algorithmically solvable turns this
classification relevant for modern computer algebra implementations too.
As shown in \se{tests}, taking as framework for instance Kamke's examples,
78\% belong to this ODE-class. Even after discarding Riccati ODEs, 75\% of
the remaining Kamke examples belong to this class \eq{GTS} and so are
solvable by the algorithm presented in \se{linear_transformations}. This
algorithm actually solves many ODE families not solved in the presently
available computer algebra systems (CAS). For instance, from the 4
examples shown in \se{examples}, three of them cannot be solved by Maple 6
or Mathematica 4 - the last versions of these major CAS. Even concerning
the Riccati families presented in \se{riccati}, two of them (cases
``$f=q$" and ``$p=q$") are also not solved by these two CAS, which base
their strategy in mapping Riccati ODEs into linear second order ones.

\medskip
\ni {\bf Acknowledgments}
\medskip

\noindent This work was supported by the Centre of Experimental and
Constructive Mathematics, Simon Fraser University, Canada, and by the
Symbolic Computation Group, Faculty of Mathematics, University of
Waterloo, Ontario, Canada. The authors would like to thank K. von
B\"ulow\footnote{Symbolic Computation Group, Faculty of Mathematics,
University of Waterloo, Ontario, Canada.} for a careful reading of this
paper.

\def\bibauthor #1{#1,}
\def\bibarticle #1{{#1}.}
\def\bibbook #1{{\it #1}.}
\def\bibsource #1{#1 }
\def\bibyear #1{#1}


\begin{thebibliography}{99}

\bibitem{bluman}
\bibauthor{Bluman G. \& Kumei S.}
\bibyear{1996}
\bibbook{Symmetries and Differential Equations}
\bibsource{New York. Springer-Verlag.}


\bibitem{ktese}
\bibauthor{von B\"ulow, K.}
\bibyear{2000}
\bibarticle{Equivalence methods for second order linear differential equations}
\bibsource{M.Sc. Thesis, Faculty of Mathematics, University of Waterloo.}

\bibitem{odetools1}
\bibauthor{Cheb-Terrab E.S., Duarte L.G.S. and da Mota L.A.C.P.}
\bibyear{1997}
\bibarticle{Computer Algebra Solving of First Order ODEs Using Symmetry Methods}
\bibsource{Computer Physics Communications, 101, 254.}

\bibitem{sym_pat}
\bibauthor{Cheb-Terrab E.S, Roche A.D.}
\bibyear{1998}
\bibarticle{Symmetries and First Order ODE patterns}
\bibsource{Computer Physics Communications 113, 239.}

\bibitem{abel} 
\bibauthor{Cheb-Terrab E.S, Roche A.D.}
\bibarticle{Abel ODEs: Equivalence and New Integrable Cases}
\bibsource{Computer Physics Communications 130, 2000, p.197.}

\bibitem{chini}
\bibauthor{Chini M.}
\bibyear{1924}
\bibarticle{Sull'integrazione di alcune equazioni differenziali del primo ordine}
\bibsource{Rendiconti Instituto Lombardo (2) {\bf 57}, 506-511.}

\bibitem{dickson}
\bibauthor{Dickson  L.E.}
\bibyear{1924}
\bibsource{Annals of Math. 2, 25, 324.}

\bibitem{hereman}
\bibauthor{Hereman W.}
\bibyear{1995}
\bibbook{Chapter 13 in vol 3 of the CRC Handbook of
Lie Group Analysis of Differential Equations}
\bibsource{Florida. Ed.: N.H.Ibragimov, CRC Press, Boca Raton.}

\bibitem{hydon}
\bibauthor{Hydon P.E.}
\bibyear{1995}
\bibarticle{Conformal symmetries of first-order ordinary differential equations}
\bibsource{J.Phys. A: Math. Gen. {\bf 27} 385-392.}

\bibitem{kamke}
\bibauthor{Kamke, E.}
\bibyear{1947}
\bibbook{Differentialgleichungen}
\bibsource{N.Y. Chelsea Publ. Co.}

\bibitem{murphy}
\bibauthor{Murphy G.M.}
\bibyear{1960}
\bibbook{Ordinary Differential Equations and  their solutions}
\bibsource{Van Nostrand, Princeton.}

\bibitem{pirondini}
\bibauthor{Pirondini  G.}
\bibyear{1904}
\bibsource{Annali di Mat. 3 (9), 185.}


\bibitem{schwarz}
\bibauthor{Schwarz  F.}
\bibyear{1998}
\bibarticle{Symmetry analysis of Abel's equation}
\bibsource{Stud. Appl. Math. 100, no. 3, 269--294.}

\bibitem{stephani}
\bibauthor{Stephani H.}
\bibyear{1989}
\bibbook{Differential equations: their solution using symmetries}
\bibsource{Cambridge University Press.}

%
%
%
%

\end{thebibliography}
\end{document}